\documentclass[12pt]{iopart}

\usepackage{graphicx}
\usepackage{caption}
\usepackage{subcaption}
\usepackage{epstopdf}
\usepackage{iopams}
\pdfoutput=1 
\usepackage{color}
\usepackage[numbers,square, comma, sort&compress]{natbib}
\usepackage{hyperref}

\newcommand{\sign}{\textup{sign}}
\newcommand\eq[1]{Eq.~(\ref{#1})}

\begin{document}
	
\title[Exact results for the temperature-field behavior of the critical Casimir force]
{Exact results for the behavior of the thermodynamic Casimir force in a model with a strong adsorption}

\author{Daniel M Dantchev, Vassil M Vassilev  \MakeLowercase{and}  Peter A Djondjorov}

\address{Institute of Mechanics, Bulgarian Academy of Sciences,
Acad. G. Bonchev St., Building 4, 1113 Sofia, Bulgaria}

\begin{abstract} 
	When massless excitations are limited or modified by the presence of material bodies one observes a force acting between them generally called Casimir force. Such excitations are present in any fluid system close to its true bulk critical point. We derive exact analytical results for both the temperature and external ordering field behavior of the thermodynamic Casimir force within the mean-field Ginzburg-Landau Ising type model of a simple fluid or binary liquid mixture. We investigate the case when under a film geometry the boundaries of the system exhibit strong adsorption onto one of the phases (components) of the system. We present analytical and numerical results for the (temperature-field) relief map of the force in both the critical region of the film close to its finite-size or bulk critical points as well as in the  capillary condensation regime below but close to the finite-size critical point.
	
	\vspace{2pc}
\end{abstract} 

\maketitle

\section{Introduction}

In a recent article \cite{DVD2015}, we have derived exact results for both the temperature $T$ and external ordering field $h$ behavior of the order parameter profile and the corresponding response functions -- local and total susceptibilities --  within the three-dimensional continuum mean-field Ginzburg-Landau Ising type model of a simple fluid or binary liquid mixture for a system with a film geometry $\infty^2 \times L$.  In the current article we extend them to derive exact results for the thermodynamic Casimir force within the same model. 
We concentrate in the region of the parametric space in $(T,h)$ plane close to the critical point of the fluid or close to the demixing point of the binary liquid mixture. We recall that for classical fluids in the case of a simple fluid or for binary liquid mixtures the wall generically prefers one of the fluid phases or one of the components. Because of that in the current article we study the case when the bounding surfaces of the system strongly prefer one of the phases of the system.  Since in such systems one observes also the phenomena of the capillary condensation close below the critical point for small negative values of the ordering field $(\beta h) (L/a)=O(1), h<0$, we also study the behavior of the force between the confining surfaces of the system in that parametric region. Here $\beta=1/(k_B T)$, the field $h$ is measured in units of Bohr magneton $\mu_B$, and $a$ is some characteristic microscopic length, say, the average distance between the constituents of the fluid.  Let us recall that the model we are going to consider is a standard model within which one studies phenomena like critical adsorption \cite{PL83, E90,FD95,THD2008,BU2001,EMT86,OO2012,MCS98,DSD2007,DMBD2009,C77,G85,TD2010,DRB2007},  wetting or drying \cite{C77,NF82,G85,BME87,Di88,SOI91}, surface phenomena \cite{Bb83,D86}, capillary condensation \cite{BME87,E90,BLM2002,EMT86,OO2012,DSD2007,YOO2013}, localization-delocalization phase transition \cite{PE90,PE92,BLM2003}, finite-size behavior of thin films \cite{KO72,NF83,FN81,BLM2003,NAFI83,EMT86,Ba83,C88,Ped90,PE90,BDT2000}, the thermodynamic Casimir effect \cite{INW86,K97,SHD2003,GaD2006,DSD2007,MGD2007,ZSRKC2007}, etc.  The results of the model have been also used to calculate the Casimir forces in systems with chemically or topographically patterned substrates, as well as, coupled with the Derjaguin approximation, for studies on interactions of colloids -- see, e.g., the review \cite{GD2011} and the literature cited therein. Until very recently, i.e. before Ref. \cite{DVD2015}, the results 
for the case $h=0$ were derived analytically \cite{INW86,K97,GaD2006,DRB2009} while the $h$-dependence was studied numerically either at the bulk critical point of the system $T=T_c$, or along some specific isotherms -- see, e.g., \cite{SHD2003,MB2005,DSD2007,DRB2007,DRB2009,PE92,LTHD2014}. In the current article we are going to improve this situation with respect to the Casimir force by deriving exact analytical results for it in the $(T,h)$ plane. 

In 1948 \cite{C48}, after a discussion with Niels Bohr \cite{C99}, the Dutch physicist H. B. G.  Casimir realized that the zero-point fluctuations of the electromagnetic field in vacuum lead to a force of attraction between two perfectly conducting parallel plates and calculated this force.  In 1978 Fisher and De Gennes \cite{FG78} pointed out that a very similar effect exists in fluids with the fluctuating field being
the field of its order parameter, in which the interactions in the system are mediated not by photons but by different type of massless excitations such as critical fluctuations or Goldstone bosons (spin waves). Nowadays one usually terms the corresponding Casimir
effect the critical or the thermodynamic Casimir effect \cite{BDT2000}. 

Currently the Casimir, and Casimir-like, effects are object of studies in
quantum electrodynamics, quantum chromodynamics, cosmology, condensed matter physics, biology and, some elements of it, in nano-technology. The interested reader can consult the existing impressive number of reviews on the subject  \cite{PMG86,MT88,LM93,MT97,M94,KG99,B99,BMM2001,M2001,M2004,L2005,KM2006,GLR2008,BKMM2009,KMM2009,FPPRJLACCGKKKLLLLWWWMHLLAOCZ2010,OGS2011,CP2011,KMM2011,RCJ2011,MAPPBE2012,B2012,Bo2012,C2012,DGT2014,RHWJLC2015,BW2007,KM2015c,WDTRRP2015,ZLP2015,K94,BDT2000,K99,G2009,TD2010,GD2011,D2012,V2015}. So far the
critical Casimir effect has enjoyed only two general reviews \cite{K94,BDT2000}
and few concerning specific aspects of it \cite{K99,G2009,TD2010,GD2011,D2012,V2015}. 

The critical Casimir effect has been already directly observed, utilizing light scattering measurements, in the interaction of a colloid spherical particle with a plate \cite{HHGDB2008} both of which are immersed in a critical binary liquid mixture. Very recently the nonadditivity of critical Casimir forces has been experimentally demonstrated in \cite{PCTBDGV2016}. Indirectly, as  a balancing force that determines the thickness of a wetting film in the vicinity of its bulk critical point the Casimir force has been also studied in $^4$He \cite{GC99}, \cite{GSGC2006}, as well as in $^3$He--$^4$He mixtures \cite{GC2002}. In  \cite{FYP2005} and   \cite{RBM2007} measurements of the Casimir force in thin wetting films of binary liquid mixture are also performed. The studies in the field have also enjoined a considerable theoretical attention.  Reviews on the corresponding results can be found in \cite{K99,G2009,TD2010,GD2011,D2012,V2015}. 

Before turning exclusively to the behavior of the Casimir force, let us briefly remind some basic facts of the theory of critical phenomena. In the vicinity of the bulk critical point  $(T_c,h=0)$ the bulk correlation length of the order parameter $\xi$ becomes large, and theoretically diverges: $\xi_t^+\equiv\xi(T\to T_c^{+},h=0)\simeq \xi_0^{+} t^{-\nu}$, $t=(T-T_c)/T_c$, and $\xi_h\equiv\xi(T=T_c,h\to 0)\simeq \xi_{0,h} |h/(k_B T_c)|^{-\nu/\Delta}$, where $\nu$ and $\Delta$ are the usual critical exponents and $\xi_0^{+}$ and $\xi_{0,h}$ are the corresponding nonuniversal amplitudes of the correlation length along the $t$ and $h$ axes.  If in a finite system  $\xi$ becomes comparable to  $L$, the thermodynamic functions describing its behavior depend on the ratio $L/\xi$ and take scaling forms given by the finite-size scaling theory. For such a system the finite-size scaling theory \cite{C88,BDT2000,Ba83,P90,Ped90,K94} predicts:

$\bullet$ For the Casimir force
\begin{equation}\label{cas}
F_{\rm Cas}(t,h,L)=L^{-d}X_{\rm Cas}(x_t,x_h);
\end{equation}

$\bullet$ For the order parameter profile
\begin{equation}\label{mfss}
\phi(z,T,h,L)= a_h L^{-\beta/\nu}X_\phi
\left(z/L,x_t,x_h\right),
\end{equation}
where
$x_t=a_t t L^{1/\nu}$, $x_h=a_h h L^{\Delta/\nu}$.
In Eqs. (\ref{cas}) and (\ref{mfss}), $\beta $ is the critical exponent for the order parameter, $d$ is the dimension of the system, $a_t$ and $a_h$ are nonuniversal metric factors that can be fixed, for a given system, by taking them to be, e.g., $a_t=1/\left[\xi_0^+\right]^{1/\nu}$, and $a_h=1/\left[\xi_{0,h}\right]^{\Delta/\nu}$. 

\section{The Ginzburg-Landau mean-field model and the Casimir force}

\subsection{Definition of the model}

Here, as in \cite{DVD2015}, we consider a critical system of Ising type in a film geometry $\infty^2\times L$, where $L$ is supposed to be along $z$ axis, described by the minimizers of the standard $\phi^4$ Ginzburg-Landau functional
\begin{equation} \label{FviafIsing}
{\cal F}\left[\phi;\tau,h,L\right] =\int_0^L {\cal L}(\phi,\phi') dz,
\end{equation}
where
\begin{equation}\label{fdefIsing}
{\cal L}\equiv {\cal L}(\phi,\phi')=\frac{1}{2}  {\phi'}^2 +
\frac{1}{2}\tau\phi^2+\frac{1}{4}g\phi^4-h \phi.
\end{equation}
Here $L$ is the film thickness, ${\phi}(z|\tau,h,L)$ is the order parameter   assumed to depend on the perpendicular position $z\in(0,L)$ only, $\tau=(T-T_c)/T_c\,(\xi_0^{+})^{-2}$ is the bare reduced temperature, $h$ is the external ordering field, $g$ is the bare coupling constant and the primes indicate differentiation with respect to the variable $z$.

\subsection{Basic expression for the Casimir force}

The thermodynamic Casimir force is the excess pressure over the bulk one acting on the boundaries of the system which is due to the finite size of the system. To derive this excess pressure there are several ways but probably the most straightforward one is to  apply the corresponding mathematical results of the variational calculus. For example, 
following Gelfand and Fomin \cite[pp. 54--56]{Gelfand1963} it is easy to show that the functional derivative of ${\cal F}$ with respect to the  independent variable $z$ at $z=L$ is
\begin{equation}
\label{GF1963}
\left.-\left(\frac{\delta \mathcal{F}}{ \delta z}\right)\right |_{z=L}=
\left.-\left(\phi' \frac{\partial \cal L}{\partial \phi'}-{\cal L} \right) \right|_{z=L}.
\end{equation}
Having in mind \eq{fdefIsing}, one derives explicitly 
\begin{eqnarray}\label{dz}
\left.-\left(\frac{\delta \mathcal{F}}{ \delta z}\right)\right|_{z=L}&=&
\left.\left(\frac{1}{2}  {\phi'}^2-\frac{1}{4}g\phi^4 - 
\frac{1}{2}\tau\phi^2+h \phi \right) \right|_{z=L}   \nonumber \\
&\equiv& p_L(\tau,h).
\end{eqnarray}
This derivative has the meaning of a force acting on the surface of the system at $z=L$ and, since ${\cal F}$ is normalized per unit area, it has a meaning of a pressure acting on that surface. That is why, the notation $p_L(\tau,h)$ is used. The above is actually the procedure used in \cite{INW86} where the authors perform the corresponding variational calculations on their own. Another common way to proceed is to use the apparatus based on the stress tensor operator (see, e.g., \cite{SHD2003},  \cite{EiS94} and \cite{VD2013})
\begin{equation}
\label{Tkl}
T_{kl} =  \frac{\partial {\cal L}}{\partial (\partial_l \Phi)} (\partial_k \Phi) -\delta_{kl}\,{\cal L}.
\end{equation}
It is elementary to check that the expression in the parentheses in the right-hand-side of Eq. \ref{dz} coincides with $T_{zz}$ component of the stress tensor. In  \cite{DVD2015}, we have shown that this expression is a first integral of the considered  system, and therefore $T_{zz}$ and $p_L(\tau,h)$ do not depend on the coordinate $z$ at which they are calculated.

In the bulk system, within the mean-field theory the gradient term in $\cal L$ is absent and instead of $p_L$ one obtains 
\begin{equation}\label{pb}
p_b(\tau,h)=-\frac{1}{4}g \phi_b^4-\frac{1}{2}\tau \phi_b^2+h \phi_b,
\end{equation}
where $\phi_b$ is the order parameter of the bulk system. Clearly, $\phi_b$ is determined by the cubic equation
$-\phi_b\left[\tau+g\,\phi_b^2\right]+h=0$, $\phi_b$ being such that
${\cal L}_b =
\frac{1}{2}\tau\phi_{b}^2+\frac{1}{4}g\phi_{b}^4-h \phi_{b}$
attains its minimum.
Now, one can immediately determine the Casimir force as
\begin{equation} \label{Casimir}
F_{\rm Cas}(\tau,h,L)= p_L(\tau,h)-p_b(\tau,h).
\end{equation}
When $F_{\rm Cas}(\tau,h,L)<0$ the excess pressure will be inward
of the system that corresponds to an {\it attraction} of the surfaces of the system towards each other and to a {\it repulsion} if $F_{\rm Cas}(\tau,h,L)>0$.

In the light of the above it is evident that once the order parameter profile $\phi$ is known in analytic form for given values of the parameters $\tau$ and $h$, then the corresponding Casimir force is determined exactly. It is noteworthy that the above expressions do not depend on the specific choice of the boundary conditions. In the current article we specialize to the so-called $(+,+)$ boundary conditions under which one requires that
$\left. \lim \phi\left( z \right) \right| _{z \rightarrow 0}=\left.
\lim \phi \left( z \right) \right| _{z \rightarrow L}=+\infty $.
The exact solution for $\phi(z,\tau,h,L)$ for this case has been determined in \cite{DVD2015}. In what follows we will study the properties of the force $F_{\rm Cas}(\tau,h,L)$ using this exact solution. 
\begin{figure}[h!]
	\centering
	\includegraphics[width=\columnwidth]{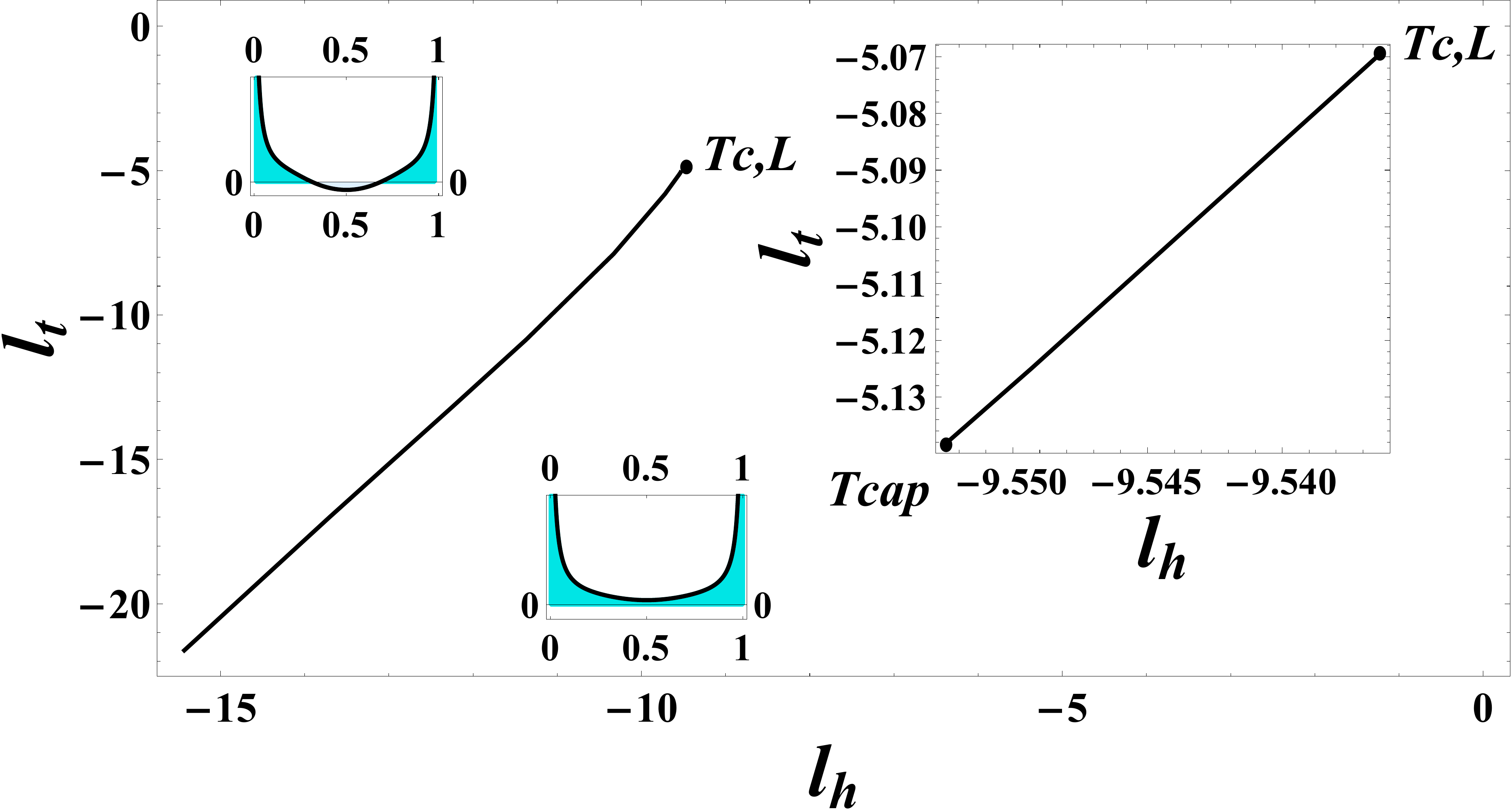}
	\caption{Phase diagram.
The critical point of the finite system is $T_{c,L} = (l_t^{(c)},l_h^{(c)})=(-5.06935, -9.53633)$. We recall that at this point the susceptibility of the finite system diverges. The inset on the right shows the pre-capillary-condensation curve, determined in \cite{DVD2015}, where above $ T_{\rm cap}=(-5.13834, -9.55252)$ and below $ T_{c,L}$ the jump of the order parameter at the middle of the system is from a less dense gas to a more dense one. For $T\le T_{\rm cap}$ the system jumps from a ''gas'' to a ''liquid'' state.} 
	\label{fig:PD}
\end{figure}

\section{Exact results for the Casimir force} 

Since the thermodynamic Casimir force is normally presented in terms of the scaling variables 
\begin{equation}\label{isvarxt}
l_t\equiv \sign(\tau) \, L/\xi_t^+ =\sign(\tau)\, L \sqrt{|\tau|},
\end{equation}
\begin{equation}\label{isvarxh}
l_h\equiv \sign(h)\, L/\xi_h=L\sqrt{3}\left(\sqrt{g}\, h\right)^{1/3}, 
\end{equation}
in the remainder we are going to use such variables as the basic parameters determining the behavior of the force. In the above we  have taken into account that for the model considered here $\xi_{0,h}/\xi_0^+=1/\sqrt{3}$ \cite{SHD2003}, $\nu=1/2$ and $\Delta=3/2$. The phase diagram of this model has been studied in details in  \cite{DVD2015} -- see there figure 3 and the text around it. Here, for the convenience of the reader, it is depicted in figure \ref{fig:PD} in terms of the scaling variables $l_t$ and $l_{h}$.

\subsection{Exact analytical results for the Casimir force}

In terms of the scaling variables given in equations (\ref{isvarxt}) and (\ref{isvarxh}), the value $p_L(\tau,h)$ of the first integral, see \eq{dz}, becomes 
\begin{equation} \label{FIX}
p_L(\tau,h)=\frac{1}{g L^4} p\left( l_t,l_h \right),
\end{equation}
where the constant $p\left( l_t,l_h \right)$ is 
\begin{equation} \label{FIc}
p\left( l_t,l_h \right)={X'}^2
-X^4-\sign(l_t) \, l_t^2 X^2 +\frac{2}{3 \sqrt{6}}l_h^{3} X.
\end{equation}
Here 
\begin{equation}\label{isvar2}
X(\zeta|l_t,l_h)=\sqrt{\frac{g}{2}}L^{\beta/\nu} \phi(z) 
\end{equation}
is the scaling function of the order parameter $\phi$,  $\beta=1/2$ and hereafter the prime means differentiation with respect to the variable $\zeta=z/L, \zeta \in [0,1]$. Similarly, for the bulk system, see \eq{pb}, one has 
\begin{equation} \label{pbX}
p_b(\tau,h)=  \frac{1}{g L^4}p_b(l_t,l_h),
\end{equation}
where
\begin{equation} \label{pbX1}
p_b(l_t,l_h)= -X_{b}^4-\sign(l_t) \, l_t^2 X_{b}^2 +\frac{2}{3 \sqrt{6}}l_h^{3} X_{b}.
\end{equation}
From Eqs. (\ref{FIX}) and (\ref{pbX}) for the Casimir force (\ref{Casimir}) one obtains 
\begin{equation} \label{CasimirX}
F_{\rm Cas}(\tau,h,L)= \frac{1}{g L^4} X_{\rm Cas}(l_t,l_h),
\end{equation}
where its scaling function $X_{\rm Cas}$ is
\begin{equation} \label{CasimirX1}
X_{\rm Cas}(l_t,l_h)= p\left( l_t,l_h \right)-p_b(l_t,l_h).
\end{equation}

Given $l_t$ and $l_h$, the determination of $p_b(l_t,l_h)$ is evident, while $p\left( l_t,l_h \right)$ is given by the expression 
\begin{equation}\label{IntConst}
p\left( l_t,l_h \right)=x_m\left( \frac{2}{3 \sqrt{6}}l_{h}^{3}-x_m^{3}-\sign(l_t) l_t^2 \, x_m\right),
\end{equation}
see Eq. (3.15) in \cite{DVD2015}. As shown in   \cite{DVD2015}, $x_m$ is to be determined from
\begin{equation}  \label{TrEq}
12 \wp \left(\frac{1}{2};g_{2},g_{3}\right) - \sign(l_t)l_t^2 -6 x_m^{2}=0
\end{equation}
so that it gives rize to a continuous order parameter profile in the interval $(0,1)$, and satisfies the condition 
\begin{equation}  \label{eq:condition}
6 \sqrt{3}\, x_m \left(\sign(l_t)l_t^2 +2 x_m^{2}\right)- \sqrt{2} \, l_{h}^{3}>0.
\end{equation}
In \eq{TrEq} $\wp \left(\xi;g_{2},g_{3}\right)$ is the Weierstrass elliptic function whose invariants  $g_{2}$ and $g_{3}$ are given by the expressions
\begin{equation} \label{g1} 
g_2=\frac{1}{12} l_t^4+p\left( l_t,l_h \right),
\end{equation}
\begin{equation} \label{g2} 
g_3=-\frac{1}{216} \left[l_{h}^6+ l_t^6-36  \, p\left( l_t,l_h \right) l_t^2\right].
\end{equation}
Thus,  in order to determine $p\left( l_t,l_h \right)$  for the regarded $(+,+)$  boundary conditions at given values of the parameters $l_t$ and $l_h$, one should find all the solutions $x_m$ of the transcendental equation (\ref{TrEq}) which meet the above requirements. If there is more than one such solution $x_ m$, as explained in detail in  \cite{DVD2015}, we take that one which leads to an order parameter profile that corresponds to the minimum of the energy functional (\ref{FviafIsing}) 
\begin{equation} \label{afunctional}
{\cal E} =\frac{1}{g L^4}\int_0^1 f(X,X') d\zeta, 
\end{equation}
where
 \begin{equation}\label{fdefIsingmScalingA}
f(X,X')=X'^2 + X^4+\sign(l_t) \, l_t^2 X^2-\frac{2}{3 \sqrt{6}}l_h^{3} X.
\end{equation}
The precise mathematical procedure how this can be achieved, despite the divergence of the energy (see Eq. (3.27) in \cite{DVD2015} and the text around it), is also explained in details in   \cite{DVD2015}. Let us note that $x_m$ has a clear physical meaning -- it is the value of the scaling function of the order parameter profile at the middle of the system, i.e., $X(1/2|l_t,l_h)=x_m(l_t,l_h)$.

From Eqs. \ref{pbX1} and \ref{IntConst}, once $X_b$ and $x_m$ are determined, the scaling function of the Casimir force takes the form
\begin{equation} \label{Casimir_final}
X_{\rm Cas}(l_t,l_h)= X_{b}^4-x_m^4+\sign(l_t) \, l_t^2 \left(X_{b}^2-x_m^2\right) -\frac{2}{3 \sqrt{6}}l_h^{3} \left(X_{b}-x_m\right).
\end{equation}

When $h=0$, i.e. $l_h=0$, the behavior of the Casimir force under (+,+) boundary conditions has been analytically studied in \cite{K97} and \cite{INW86}. In   \cite{INW86} the value of the so-called Casimir amplitude, i.e., the result for $l_t=l_h=0$ is obtained, while in \cite{K97} the behavior of the force as a function of $l_t$ has been studied. 

When $h \neq 0$ the behavior of the Casimir force has been studied only numerically. In Refs. \cite{SHD2003} and \cite{VD2013} it has been obtained only for $T=T_c$ for some chosen values of $l_h$. Below we present its behavior as a function of both $l_t$ and $l_h$ in $(l_t,l_h)$ plane  by evaluating numerically the analytical expressions given above. 

\subsection{Numerical evaluation of the analytical expressions}
\label{sec:num_results}
\begin{figure}[h!]
\centering
\includegraphics[width=4.4in]{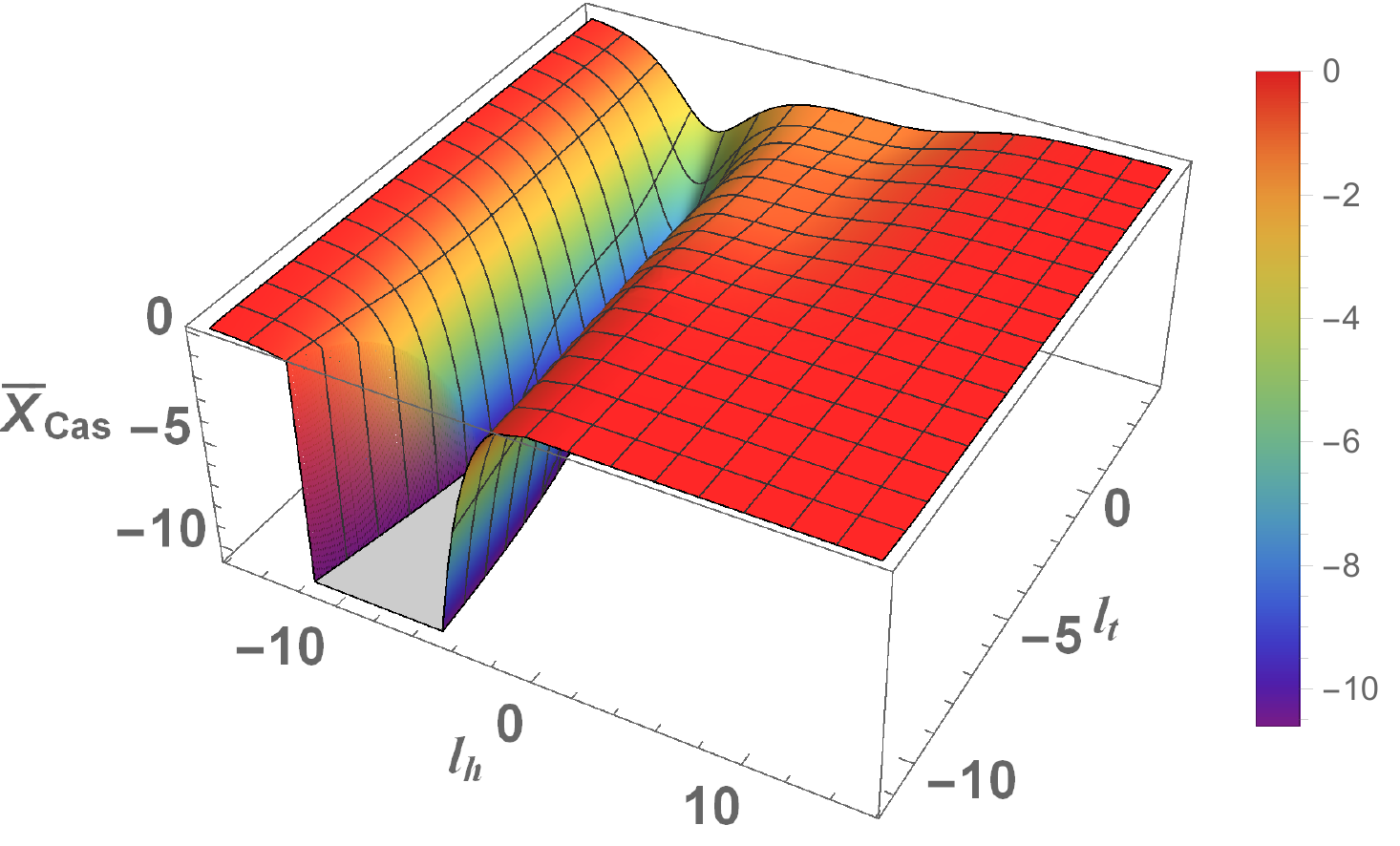} \quad
\includegraphics[width=4.4in]{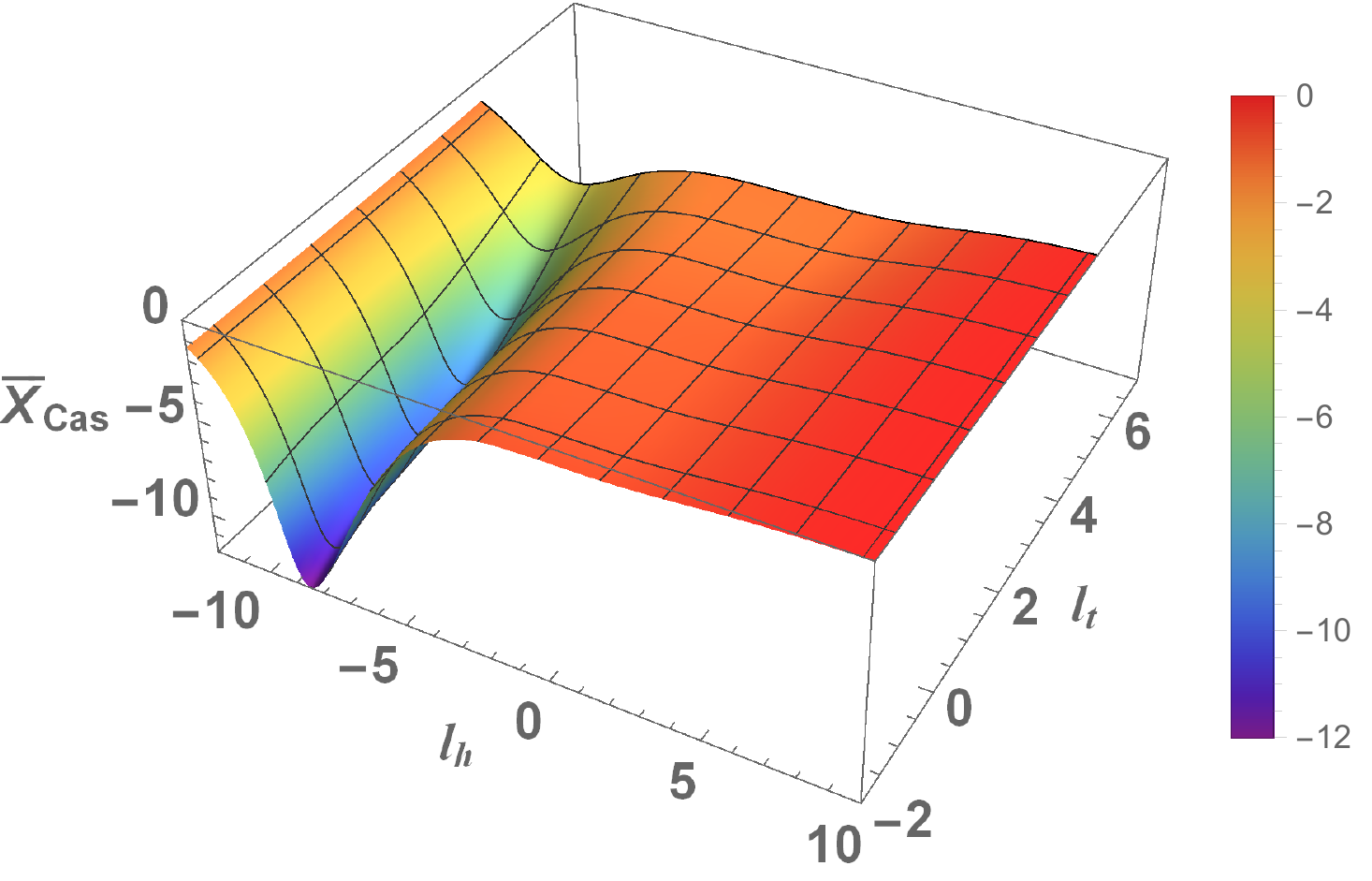}
\caption{Plots of Casimir force as a function of both $l_t$ and $l_h$. }
\label{fig:DP}
\end{figure}

\begin{figure}
    \centering
    \begin{subfigure}[h!]{0.775\textwidth}
        \includegraphics[width=\textwidth]{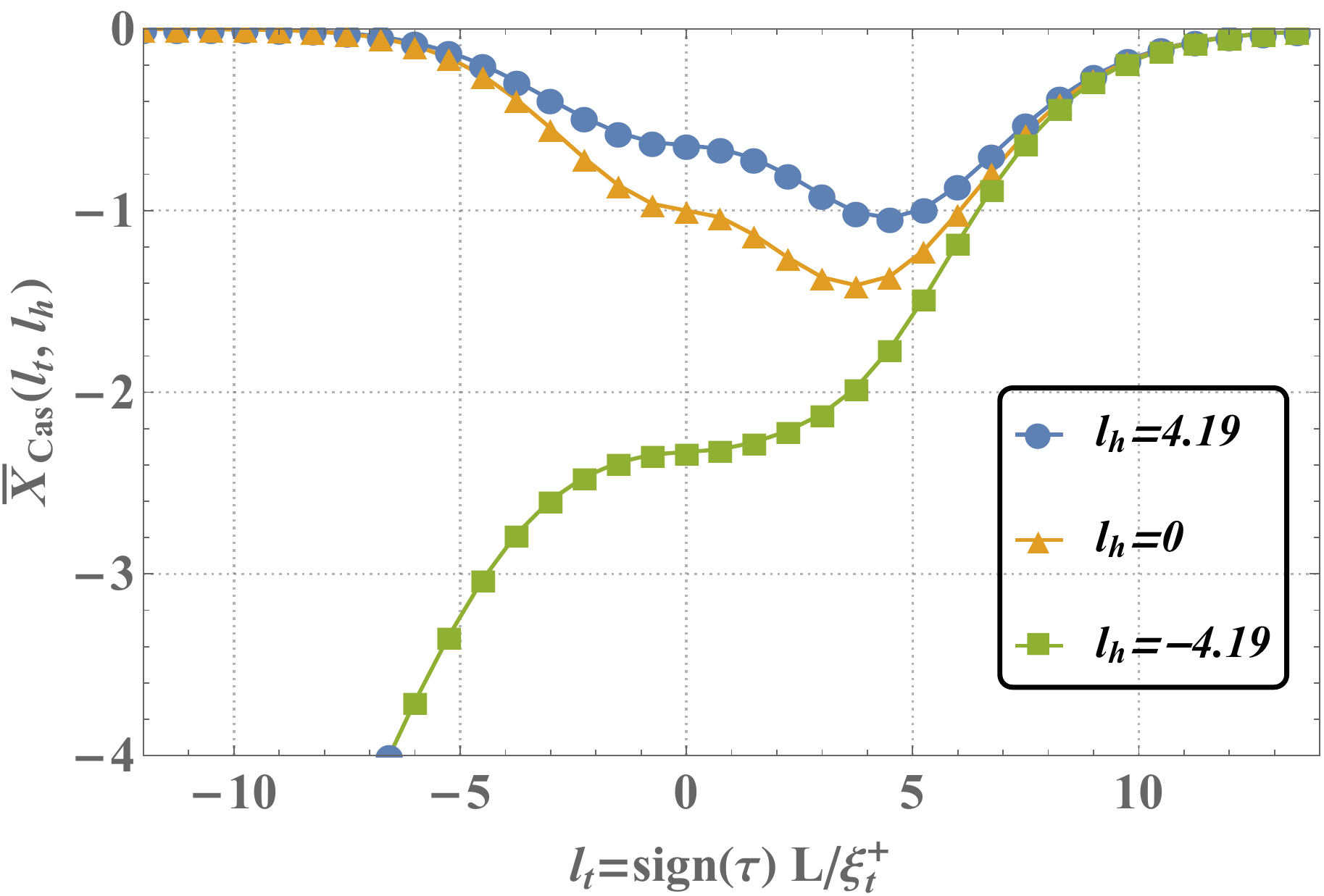}
        \caption{The dependence of the normalized finite-size scaling function $\overline{X}_{\rm Cas} (l_t,l_h) $ of the Casimir force on the scaling variable $l_t$ for three values of the scaling variable $l_h$: $l_h=0$, $l_h=\pm 4.19$. }
        \label{CasPlusMF}
    \end{subfigure}
    
    \quad 

    \begin{subfigure}[h!]{0.765\textwidth}
        \includegraphics[width=\textwidth]{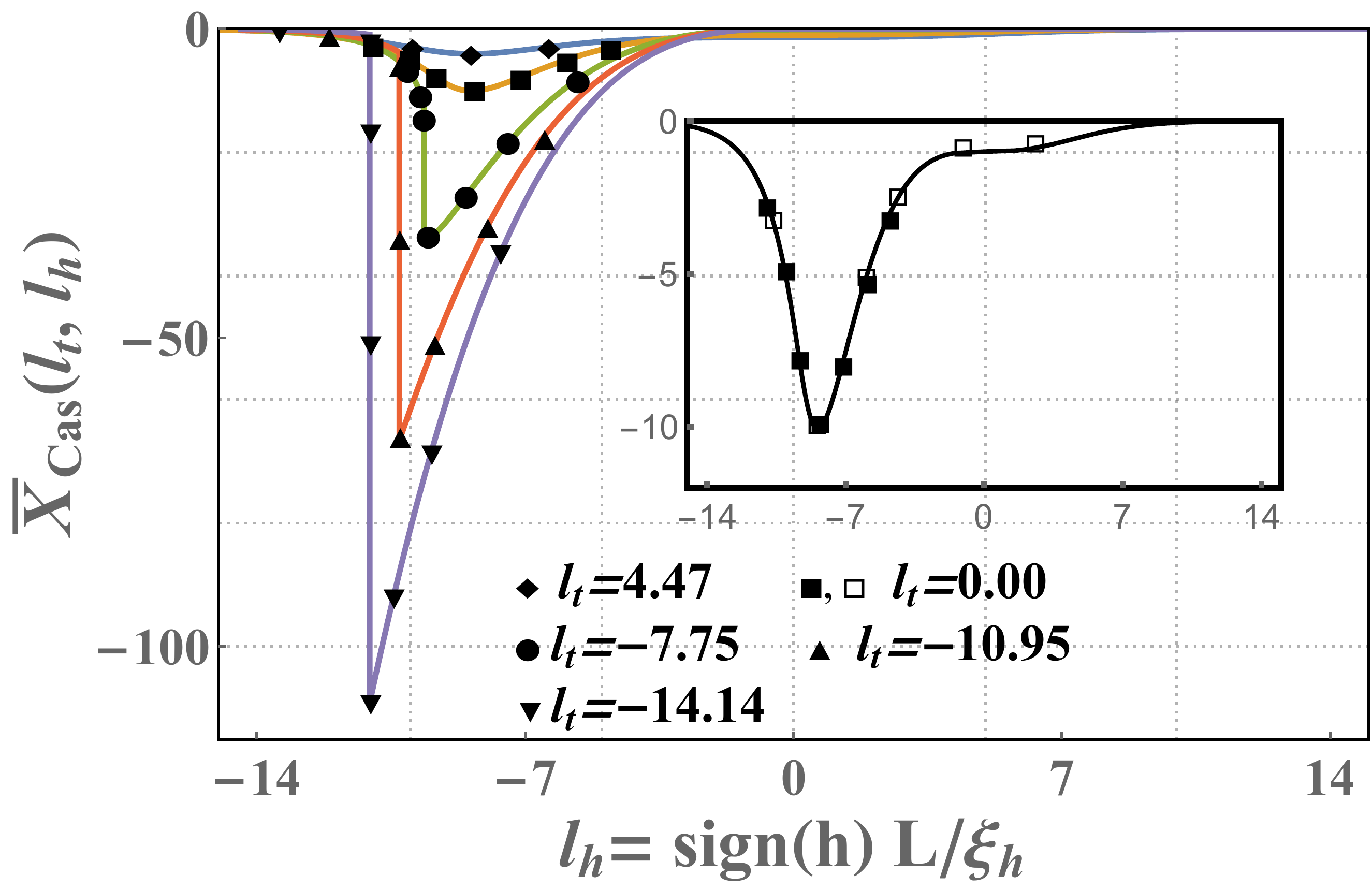}
        \caption{The dependence of the normalized thermodynamic Casimir force $\overline{X}_{\rm Cas} (l_t,l_h)$ on the field scaling variable $l_h$ for several values of the temperature scaling variable $l_t$. }
         \label{fig:CF}
    \end{subfigure}
        \caption{Plots of cross-sections of the Casimir force for given fixed values of $l_t$, or $l_h$, as a function of  $l_h$, or $l_t$, respectively.}\label{fig:CS}
\end{figure}

\begin{figure}[h!]
	\centering
	\includegraphics[width=0.705\textwidth]{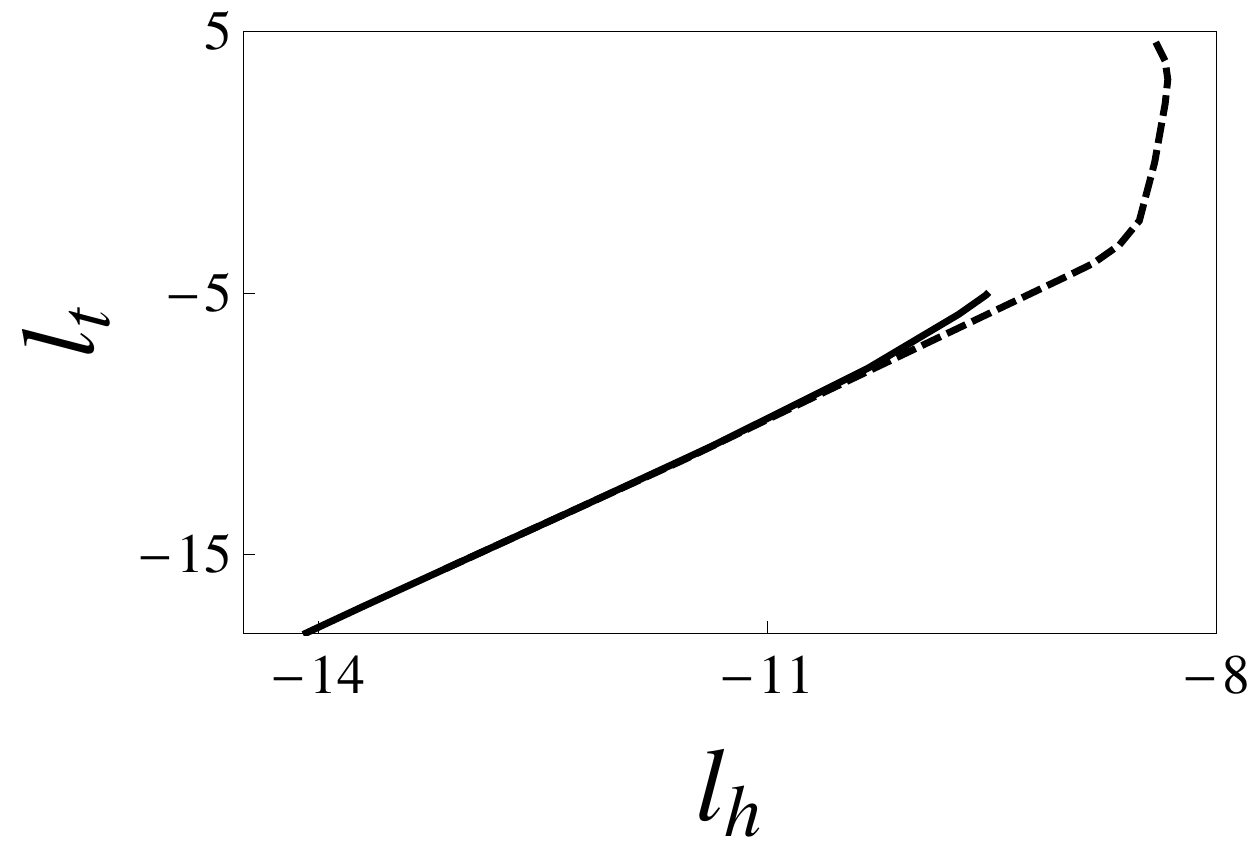}
	\caption{The phase digram curve (solid; $ - 15.45 \leq l_t \leq - 5.07$) vs the curve (dashed; $-15.49 \leq l_t \leq 4.47$ ) in the $(l_t,l_h)$ plane at which the Casimir force attains minimum for a given temperature $l_t$. Note that for relatively small negative values of $l_t$, the jump in the force when one  crosses the phase coexistence curve and the minimal Casimir force actually occur at different values of the field $l_h$.  The last implies that by increasing the temperature one passes though the minimum of the force which then continuously increases till crossing the coexistence curve at which the force jumps to negative values much closer to $0$. For $l_t\lesssim -10$ the minimal value of the force is practically achieved at the phase coexisting line.}
	\label{fig:PhD_vs_MCF}
\end{figure}

Using the derived exact analytical expressions described above in the current section we determine the Casimir force in the critical and in the capillary condensation regimes at temperatures below but relatively close to $T_{c,L}$. It should be pointed out that the solutions $x_m$ of the transcendental equation (\ref{TrEq}) that correspond to certain values of the parameters $l_t$ and $l_{h}$ are to be obtained numerically identifying  by inspection those of them that obey the conditions formulated above.

The behavior of the normalized finite-size scaling function $\overline{X}_{\rm Cas}(l_t,l_h) \equiv X_{\rm Cas}(l_t,l_h)/|X_{\rm Cas}(0, 0)|$  of the Casimir force is shown in figures \ref{fig:DP}
and \ref{fig:CS}. 

The relief map of the Casimir force, as a function of both $l_t$ and $l_h$, is shown in figure \ref{fig:DP} where the upper part presents the force in a larger scale, while the lower one is a blow up of the region close to the bulk critical point. The only other model we are aware of where such a relief map as a function of both relevant scaling variables is available is that one of the three-dimensional spherical model under periodic boundary conditions \cite{D96,D98}. One observes a valley in this map with its deepest point at the dashed line shown in figure 	\ref{fig:PhD_vs_MCF}. 

Figures \ref{CasPlusMF} and \ref{fig:CF} present cross-sections of the foregoing 3d figures for given fixed values of $l_t$, or $l_h$, as a function of  $l_h$, or $l_t$, respectively. Please note that they cover region of parameters that goes beyond the one covered in figure \ref{fig:DP}. Figure \ref{CasPlusMF} shows the behavior of $\overline{X}_{\rm Cas}$ as a function of $l_t$ for $l_h=0,\pm 4.19$. Note that $\overline{X}_{\rm Cas}$ is negative
and for $l_h=0$ has a minimum at $l_t=3.749$ {\it  above} $T_c$, as in the case of the $2d$ Ising model \cite{ES94}. The value of the minimum is $1.411$ times deeper than the corresponding value of the force at $T=T_c$, which agrees with the results of \cite{K97,HSED98}. The overall behavior of the curves is similar to that one obtained for the $d=2$ Ising model \cite{ZMD2013} via density-matrix renormalization group (DMRG) calculations.  Let us, nevertheless, note that  for bulk $d=2$ Ising type systems there is no true phase transition for finite $L$. It has been shown, however, that something similar does exist -- a line
of very weakly rounded first-order transitions ending in a pseudo-critical point \cite{PF83}. Let us note that for $T_{c,L}<T<T_c$, i.e., $l_t^{(c)}<l_t<0$, there is a special region of values of the field parameter when $l_h^{(c)}<l_h<0$. For such values of $l_h$ one does not cross the capillary condensation line when lowering the temperature -- see figure \ref{fig:PD}. 
Let us also note that, of course, one formally can consider temperatures well below $T_{c,L}$ but then the model will no longer deliver physically sensible information. We recall that in the current study when taking $T<T_{c,L}$ we do consider temperatures below but quite close to $T_{c,L}$. 

Figure \ref{fig:CF}  depicts the behavior of $\overline{X}_{\rm Cas}$ as a function of $l_h$ for $l_t=4.47,\, 0, -7.75, -10.95, -14.14$. Note that the minimum of the function $\overline{X}_{\rm Cas}(0,l_h)$ is again negative, it is attained at $l_h=-8.405$, and is $10.052$ times deeper than the corresponding value at the bulk critical point. The markers on the curves, including the inset curve representing the blow-up in the case $l_t=0$, show an excellent agreement of the numerical results obtained in   \cite{SHD2003} (filled markers), and in   \cite{VD2013} (empty squares) with the analytic results (solid lines) presented here. We observe that the Casimir force exhibits
a discontinuous jump on crossing the coexistence line as predicted in \cite{E90book,EMT86} and as also shown in \cite{DME2000,SHD2003,DSD2007}. It is easy to estimate the magnitude of this jump. One way of arguing is through direct formal use of the Kelvin equation \cite{EM87} -- as it is done in \cite{DME2000}, which leads to the conclusion that the jump $\Delta F_{\rm Cas}\simeq 2 \Phi^*_b(T)h_{\rm cap}(T)\simeq -2\sigma(T)/L$ where $\Phi^*_b(T)$ is the bulk spontaneous order parameter,  $h_{\rm cap}(T)$ is the field on the capillary coexistence line and $\sigma(T)$ is the interfacial tension between the coexisting $"+"$ and $"-"$~phases of the fluid. Here we have taken into account that according to the Kelvin equation for a fixed $T<T_{c,L}$ and large $L$ coexistence happen at $h_{\rm cap}\sim \sigma(T)/[L\Phi^*_b(T)]$ \cite{EM87,DME2000}. The above expression for $\Delta F_{\rm Cas}$ implies that the
magnitude of the jump decreases in the same
fashion as the interfacial tension, as $T$ increases at fixed
$L$ towards $T_{c,L}$. Since the jumps is to an exponentially small force upon crossing the capillary condensation line the above estimate for the jump of the force can be also considered as an very rough estimation of the maximal value of the force for a given $T$ near the line of coexistence of the two phases. Let us also note that for $T$ below but close to $T_{c,L}$ the above expressions for $\Delta F_{\rm Cas}$ take a scaling form:  $2 \Phi^*_b(T)h_{\rm cap}(T)\sim |t|^\beta h\sim x_t^\beta x_h L^{-d}$ and $\sigma(T)/L\sim |t|^{(d-1)\nu}/L\sim x_t^{(d-1)}L^{-d}$ , as it is actually also clear from the scaling relation we have derived for that region of thermodynamic parameters. We recall that within the mean-field approach $\beta=\nu=1/2$ and, formally, $d=4$ in the scaling relations. We see that, as argued in \cite{DME2000}, in a fluid confined by identical, strongly adsorbing walls the Casimir force is strongly influenced by
capillary condensation. The DMRG results presented there (see figure 1 in \cite{DME2000}) and in \cite{DMC2000} for $d=2$ Ising strip subject to identical surface fields also support this. Let us note that  for $d=2$ at $T=T_c$ the scaling function of the Casimir force has a minimum  about 100 times bigger than the Casimir amplitude \cite{DMC2000}.

We conclude that our results given in Figures \ref{fig:DP}, \ref{CasPlusMF} and \ref{fig:CF} are in full conformity with the statement made in \cite{DME2000} that in a fluid confined by identical, strongly adsorbing walls the Casimir force is much stronger for states which lie slightly off bulk coexistence, with $h<0$.

\section{Summary and concluding remarks}

We have derived exact analytical results for the thermodynamic  Casimir force, see \eq{Casimir_final}, in a widely used model in the theory of phase transitions. In this model, the value $x_m$ of the order parameter in the middle of the system is a solution of \eq{TrEq} obeying the condition (\ref{eq:condition}). If there is more than one solution $x_ m$ satisfying the above requirements, we take the one that leads to an order parameter profile  corresponding to the minimum of the energy functional (\ref{afunctional}), as explained in details in   \cite{DVD2015}. The obtained results allow us to plot the relief map of the force, see figure \ref{fig:DP}, as a function of the both relevant scaling variables -- the temperature and field. In addition, figures \ref{CasPlusMF} and \ref{fig:CF} present cross-sections of the behavior of the force for given fixed values of $l_t$, or $l_h$, as a function of  $l_h$, or $l_t$, respectively. Finally, figure \ref{fig:PhD_vs_MCF} presents the loci of the minima of the force in the $(l_t,l_h)$ plane. The inspection of the results convincingly demonstrates that for $h<0$ the Casimir force in the capillary condensation regime is much more attractive than that found at the critical point.  The comparison there of the numerical evaluation of our analytical expressions with the available numerical results reported in \cite{SHD2003,VD2013} shows an excellent agreement between each other. Furthermore, the analysis of the available DMRG numerical results \cite{DME2000,ZMD2013,DMC2000} for the behavior of the Casimir force for the 2d Ising model, as presented in Sect. \ref{sec:num_results}, led us to conclusion that our mean-field results, which formally correspond to systems with spacial dimension $d\geq 4$, are quite similar in the capillary condensation regime to those of the two-dimensional Ising model. Thus, the results for the capillary condensation regime are quite robust with respect to the influence of the dimensionality. Of course, the situation is different in the parametric space close to the bulk critical point, as well as very close to the critical point of the finite system. We recall that this later critical point shall show the singularities of the corresponding $(d-1)$-dimensional system. Within the mean-field approach, as it is well known, the role of the fluctuations is not properly taken into account. If one wants to go beyond the mean-field approximation for, say, $d=3$ dimensional systems one shall either relay on numerical methods like in \cite{VD2013,V2014}, or use methods based on renormalization group approach.  However, let us recall that the mean-field results serve as a starting point there for renormalization group calculations \cite{D86,K97,P90}. Thus, our results shall be  helpful for such future analytical studies on the thermodynamic Casimir force. Finally, let us also
remind that in physical chemistry and, more precisely, in colloid sciences fluid mediated interactions
between two surfaces or large particles are usually referred to as solvation forces or disjoining pressure \cite{E90book,E90,ES94}. Thus, our results can be also considered as pertaining to a
particular case of such forces when the fluid is near its critical
point.

\providecommand{\newblock}{}


\begin{thebibliography}{100}
	\expandafter\ifx\csname url\endcsname\relax
	\def\url#1{{\tt #1}}\fi
	\expandafter\ifx\csname urlprefix\endcsname\relax\def\urlprefix{URL }\fi
	\providecommand{\eprint}[2][]{\url{#2}}


\bibitem{DVD2015}
Dantchev D~M, Vassilev V~M and Djondjorov P~A 2015 {\em Journal of Statistical Mechanics: Theory and Experiment\/} {\bf 2015} P08025

\bibitem{PL83}
Peliti L and Leibler S 1983 {\em Journal of Physics C: Solid State Physics\/}
  {\bf 16} 2635 

\bibitem{E90}
Evans R 1990 {\em J. Phys.: Condens. Matter\/} {\bf 2} 8989

\bibitem{FD95}
Fl\"{o}ter G and Dietrich S 1995 {\em Zeitschrift f\"{u}r Physik B Condensed Matter\/} {\bf 97} 213--232

\bibitem{THD2008}
Tr\"ondle M, Harnau L and Dietrich S 2008 {\em J. Chem. Phys.\/} {\bf 129} 124716

\bibitem{BU2001}
Borjan Z and Upton P~J 2001 {\em Phys. Rev. E\/} {\bf 63} 065102

\bibitem{EMT86}
Evans R, Marconi U~M~B and Tarazona P 1986 {\em J. Chem. Phys.\/} {\bf 84}  2376--2399

\bibitem{OO2012}
Okamoto R and Onuki A 2012 {\em The Journal of Chemical Physics\/} {\bf 136} 114704

\bibitem{MCS98}
Maci{\`o}{\l}ek A, Ciach A and Stecki J 1998 {\em J. Chem. Phys.\/} {\bf 108} 5913--5921

\bibitem{DSD2007}
Dantchev D, Schlesener F and Dietrich S 2007 {\em Phys. Rev. E\/} {\bf 76}  011121

\bibitem{DMBD2009}
Drzewi\ifmmode~\acute{n}\else \'{n}\fi{}ski A, Macio\l{}ek A,
  Barasi\ifmmode~\acute{n}\else \'{n}\fi{}ski A and Dietrich S 2009 {\em Phys. Rev. E\/} {\bf 79}(4) 041145


\bibitem{C77}
Cahn J~W 1977 {\em The Journal of Chemical Physics\/} {\bf 66} 3667--3672

\bibitem{G85}
de~Gennes P~G 1985 {\em Rev. Mod. Phys.\/} {\bf 57}(3) 827--863

\bibitem{TD2010}
Toldin F~P and Dietrich S 2010 {\em J. Stat. Mech\/} {\bf 11} P11003

\bibitem{DRB2007}
Dantchev D, Rudnick J and Barmatz M 2007 {\em Phys. Rev. E\/} {\bf 75} 011121

\bibitem{NF82}
Nakanishi H and Fisher M~E 1982 {\em Phys. Rev. Lett.\/} {\bf 49} 1565--1568

\bibitem{BME87}
Bruno E, Marconi U~M~B and Evans R 1987 {\em Physica 141A\/}  187--210

\bibitem{Di88}
Dietrich S 1988 Wetting phenomena {\em Phase Transitions and Critical
  Phenomena\/} vol~12 ed Domb C and Lebowitz J~L (Academic, New York) p~1

\bibitem{SOI91}
Swift M~R, Owczarek A~L and Indekeu J~O 1991 {\em EPL (Europhysics Letters)\/} {\bf 14} 475 -- 481

\bibitem{Bb83}
Binder K 1983 {\em Phase Transitions and Critical Phenomena\/} vol~8 (Academic,  London) chap~1, pp 1--144

\bibitem{D86}
Diehl H~W 1986 Field-theoretical approach to critical behavior of surfaces {\em   Phase Transitions and Critical Phenomena\/} vol~10 ed Domb C and Lebowitz J~L   (Academic, New York) p~76

\bibitem{BLM2002}
Binder K, Landau D and M\"{u}ller M 2003 {\em Journal of Statistical Physics\/} {\bf 110} 1411--1514

\bibitem{YOO2013}
Yabunaka S, Okamoto R and Onuki A 2013 {\em Phys. Rev. E\/} {\bf 87}(3) 032405

\bibitem{PE90}
Parry A~O and Evans R 1990 {\em Phys. Rev. Lett.\/} {\bf 64} 439--442

\bibitem{PE92}
Parry A~O and Evans R 1992 {\em Physica A\/} {\bf 181} 250

\bibitem{BLM2003}
Binder K, Landau D and M\"{u}ller M 2003 {\em Journal of Statistical Physics\/} {\bf 110}(3) 1411--1514

\bibitem{KO72}
Kaganov M~I and Omel`yanchuk A~N 1972 {\em JETP\/} {\bf 34} 895--898

\bibitem{NF83}
Nakanishi H and Fisher M~E 1983 {\em J. Chem. Phys.\/} {\bf 78} 3279--3293

\bibitem{FN81}
Fisher M~E and Nakanishi H 1981 {\em J. Chem. Phys.\/} {\bf 75} 5857--5863

\bibitem{NAFI83}
Nakanishi H and Fisher M~E 1983 {\em J. Phys. C: Solid State Phys.\/} {\bf 16} L95--L97

\bibitem{Ba83}
Barber M~N 1983 Finite-size scaling {\em Phase Transitions and Critical
  Phenomena\/} vol~8 ed Domb C and Lebowitz J~L (Academic, London) p 145

\bibitem{C88}
Cardy J~L (ed) 1988 {\em Finite-Size Scaling\/} (North-Holland)

\bibitem{Ped90}
Privman V (ed) 1990 {\em Finite Size Scaling and Numerical Simulation of
  Statistical Systems\/} (World Scientific, Singapore)

\bibitem{BDT2000}
Brankov J~G, Dantchev D~M and Tonchev N~S 2000 {\em The Theory of Critical Phenomena in Finite-Size Systems -- Scaling and Quantum Effects\/} (World Scientific, Singapore)

\bibitem{INW86}
Indekeu J~O, Nightingale M~P and Wang W~V 1986 {\em Phys. Rev. B\/} {\bf 34}(1) 330--342

\bibitem{K97}
Krech M 1997 {\em Phys. Rev. E\/} {\bf 56} 1642--1659

\bibitem{SHD2003}
Schlesener F, Hanke A and Dietrich S 2003 {\em J. Stat. Phys.\/} {\bf 110} 981

\bibitem{GaD2006}
Gambassi A and Dietrich S 2006 {\em J. Stat. Phys.\/} {\bf 123} 929

\bibitem{MGD2007}
Maci{\`o}{\l}ek A, Gambassi A and Dietrich S 2007 {\em Phys. Rev. E\/} {\bf 76} 031124

\bibitem{ZSRKC2007}
Zandi R, Shackell A, Rudnick J, Kardar M and Chayes L~P 2007 {\em Phys. Rev. E\/} {\bf 76} 030601

\bibitem{GD2011}
Gambassi A and Dietrich S 2011 {\em Soft Matter\/} {\bf 7} 1247--1253

\bibitem{DRB2009}
Dantchev D, Rudnick J and Barmatz M 2009 {\em Phys. Rev. E\/} {\bf 80}(3) 031119

\bibitem{MB2005}
M\"{u}ller M and Binder K 2005 {\em Journal of Physics: Condensed Matter\/} {\bf 17} S333

\bibitem{LTHD2014}
{Labb{\'e}-Laurent} M, {Tr{\"o}ndle} M, {Harnau} L and {Dietrich} S 2014 {\em Soft Matter\/}  2270

\bibitem{C48}
Casimir H~B 1948 {\em Proc. K. Ned. Akad. Wet.\/} {\bf 51} 793

\bibitem{C99}
Casimir H~B~G 1999 Some remarks on the history of the so called casimir effect   {\em The Casimir Effect 50 Years Later\/} Proceedings of the Fourth Workshop   on Quantum Field Theory under the Influence of External Conditions, Leipzig,   1998 ed Bordag M (World Scientific) pp 3--9

\bibitem{FG78}
Fisher M~E and de~Gennes P~G 1978 {\em C. R. Seances Acad. Sci. Paris Ser. B\/} {\bf 287} 207

\bibitem{PMG86}
Plunien G, M\"{u}ller B and Greiner W 1986 {\em Physics Reports\/} {\bf 134} 87--193

\bibitem{MT88}
Mostepanenko V~M and Trunov N~N 1988 {\em Soviet Physics Uspekhi\/} {\bf 31} 965

\bibitem{LM93}
Levin F~S and Micha D~A (eds) 1993 {\em Long-Range Casimir Forces\/} Finite Systems and Multiparticle Dynamics (Springer, Berlin)

\bibitem{MT97}
Mostepanenko V~M and Trunov N~N 1997 {\em The Casimir Effect and its Applications\/} (Energoatomizdat, Moscow, 1990, in Russian; English version: Clarendon, New York)

\bibitem{M94}
Milonni P~W 1994 {\em The Quantum Vacuum\/} (Academic, San Diego)

\bibitem{KG99}
Kardar M and Golestanian R 1999 {\em Rev. Mod. Phys.\/} {\bf 71} 1233--1245

\bibitem{B99}
Bordag M (ed) 1999 {\em The Casimir Effect 50 Years Later\/} (World Scientific)
  proceedings of the Fourth Workshop on Quantum Field Theory under the
  Influence of External Conditions

\bibitem{BMM2001}
Bordag M, Mohideen U and Mostepanenko V~M 2001 {\em Phys. Rep.\/} {\bf 353}
  1–205

\bibitem{M2001}
Milton K~A 2001 {\em The Casimir Effect: Physical Manifestations of Zero-point
  Energy\/} (World Scientific, Singapore)

\bibitem{M2004}
Milton K~A 2004 {\em J. Phys. A: Math. Gen.\/} {\bf 37} R209--–R277

\bibitem{L2005}
Lamoreaux S~K 2005 {\em Rep. Prog. Phys.\/} {\bf 68} 201–236

\bibitem{KM2006}
Klimchitskaya G~L and Mostepanenko V~M 2006 {\em Contemporary Physics\/} {\bf  47} 131--144

\bibitem{GLR2008}
Genet C, Lambrecht A and Reynaud S 2008 {\em Eur. Phys. J. Special Topics\/} {\bf 160} 183–193

\bibitem{BKMM2009}
Bordag M, Klimchitskaya G~L, Mohideen U and Mostepanenko V~M 2009 {\em Advances in the Casimir effect\/} (Oxford University Press, Oxford)

\bibitem{KMM2009}
Klimchitskaya G~L, Mohideen U and Mostepanenko V~M 2009 {\em Rev. Mod. Phys.\/} {\bf 81}(4) 1827--1885

\bibitem{FPPRJLACCGKKKLLLLWWWMHLLAOCZ2010}
French R~H, Parsegian V~A, Podgornik R, Rajter R~F, Jagota A, Luo J, Asthagiri D, Chaudhury M~K, Chiang Y~m, Granick S, Kalinin S, Kardar M, Kjellander R,  Langreth D~C, Lewis J, Lustig S, Wesolowski D, Wettlaufer J~S, Ching W~Y, Finnis M, Houlihan F, von Lilienfeld O~A, van Oss C~J and Zemb T 2010 {\em Rev. Mod. Phys.\/} {\bf 82}(2) 1887--1944

\bibitem{OGS2011}
Sergey D~O, Sáez-Gómez D and Xambó-Descamps S (eds) 2011 {\em Cosmology, Quantum Vacuum and Zeta Functions\/} ({\em Springer Proceedings in Physics\/} no 137) (Springer, Berlin)

\bibitem{CP2011}
Dalvit D, Milonni P, Roberts D and da~Rosa F~E (eds) 2011 {\em Casimir
  Physics\/} 1st ed ({\em Lecture Notes in Physics\/} vol 834) (Springer,
  Berlin)

\bibitem{KMM2011}
Klimchitskaya G~L, Mohideen U and Mostepanenko V~M 2011 {\em Int. J. Mod. Phys. B\/} {\bf 25} 171--230

\bibitem{RCJ2011}
Rodriguez A~W, Capasso F and Johnson S~G 2011 {\em Nature Photonics\/} {\bf 5} 211--221

\bibitem{MAPPBE2012}
Milton K~A, Abalo E~K, Parashar P, Pourtolami N, Brevik I and Ellingsen S~A  2012 {\em J. Phys. A: Math. Gen.\/} {\bf 45} 374006

\bibitem{B2012}
Brevik I 2012 {\em Journal of Physics A: Mathematical and Theoretical\/} {\bf 45} 374003

\bibitem{Bo2012}
{Bordag} M 2012 {\em ArXiv e-prints\/} (\textit{Preprint} \eprint{1212.0213})

\bibitem{C2012}
Cugnon J 2012 {\em Few-Body Systems\/} {\bf 53} 181--188

\bibitem{DGT2014}
Robert A~D~J, Vivekanand V~G and Alexandre T 2014 {\em Journal of Physics: Condensed Matter\/} {\bf 26} 213202

\bibitem{RHWJLC2015}
Rodriguez A~W, Hui P~C, Woolf D~P, Johnson S~G, Lon\v{c}ar M and Capasso F 2015 {\em Annalen der Physik\/} {\bf 527} 45--80

\bibitem{BW2007}
Buhmann S~Y and Welsch D~G 2007 {\em Progress in Quantum Electronics\/} {\bf 31} 51--130

\bibitem{KM2015c}
{Klimchitskaya} G~L and {Mostepanenko} V~M 2015 {\em Proceedings of Peter the Great St. Petersburg Polytechnic University\/} {\bf N1 (517)} 41--65

\bibitem{WDTRRP2015}
{Woods} L~M, {Dalvit} D~A~R, {Tkatchenko} A, {Rodriguez-Lopez} P, {Rodriguez}  A~W and {Podgornik} R 2015 {\em ArXiv e-prints\/} (\textit{Preprint} \eprint{1509.03338})

\bibitem{ZLP2015}
Zhao R, Luo Y and Pendry J 2015 {\em Science Bulletin\/} 1--9

\bibitem{K94}
Krech M 1994 {\em Casimir Effect in Critical Systems\/} (World Scientific,
  Singapore)

\bibitem{K99}
Krech M 1999 {\em J. Phys.: Condens. Matter\/} {\bf 11} R391

\bibitem{G2009}
Gambassi A 2009 {\em J. Phys.: Conf. Ser.\/} {\bf 161} 012037

\bibitem{D2012}
Dean D~S 2012 {\em Physica Scripta\/} {\bf 86} 058502

\bibitem{V2015}
Vasilyev O~A 2015 {\em Monte Carlo Simulation of Critical Casimir Forces\/} (World Scientific) chap~2, pp 55--110

\bibitem{HHGDB2008}
Hertlein C, Helden L, Gambassi A, Dietrich S and Bechinger C 2008 {\em
  Nature\/} {\bf 451} 172--175

\bibitem{PCTBDGV2016}
Paladugu S, Callegari A, Tuna Y, Barth L, Dietrich S, Gambassi A and Volpe G  2016 {\em Nat Commun\/} {\bf 7} 11403

\bibitem{GC99}
Garcia R and Chan M~H~W 1999 {\em Phys. Rev. Lett.\/} {\bf 83} 1187--1190

\bibitem{GSGC2006}
Ganshin A, Scheidemantel S, Garcia R and Chan M~H~W 2006 {\em Phys. Rev. Lett.\/} {\bf 97} 075301

\bibitem{GC2002}
Garcia R and Chan M~H~W 2002 {\em Phys. Rev. Lett.\/} {\bf 88} 086101

\bibitem{FYP2005}
Fukuto M, Yano Y~F and Pershan P~S 2005 {\em Phys. Rev. Lett.\/} {\bf 94} 135702

\bibitem{RBM2007}
Rafa\"\i S, Bonn D and Meunier J 2007 {\em Physica A\/} {\bf 386} 31

\bibitem{P90}
Privman V 1990 {\em Finite Size Scaling and Numerical Simulations of
  Statistical Systems\/} (World Scientific, Singapore) chap Finite-size scaling
  theory, p~1

\bibitem{Gelfand1963}
Gelfand I~M and Fomin S~V 1963 {\em Calculus of variations\/} revised english edition translated and edited by Richard A. Silverman (Prentice-Hall Inc., Englewood Cliffs, NJ)

\bibitem{EiS94}
Eisenriegler E and Stapper M 1994 {\em Phys. Rev. B\/} {\bf 50} 10009--10026

\bibitem{VD2013}
Vasilyev O~A and Dietrich S 2013 {\em EPL (Europhysics Letters)\/} {\bf 104} 60002

\bibitem{D96}
Dantchev D 1996 {\em Phys. Rev. E\/} {\bf 53} 2104--2109

\bibitem{D98}
Dantchev D~M 1998 {\em Phys. Rev. E\/} {\bf 58} 1455--1462

\bibitem{ES94}
Evans R and Stecki J 1994 {\em Phys. Rev. B\/} {\bf 49} 8842--8851

\bibitem{HSED98}
Hanke A, Schlesener F, Eisenriegler E and Dietrich S 1998 {\em Phys. Rev.
  Lett.\/} {\bf 81} 1885--1888

\bibitem{ZMD2013}
Zubaszewska M, Macio\l{}ek A and Drzewi\ifmmode~\acute{n}\else \'{n}\fi{}ski A  2013 {\em Phys. Rev. E\/} {\bf 88}(5) 052129

\bibitem{PF83}
Privman V and Fisher M~E 1983 {\em J. Phys. A: Math. Gen.\/} {\bf 16}
L295--L301

\bibitem{E90book}
Evans R 1990 {\em Liquids at Interfaces\/} (Elsevier, Amsterdam)

\bibitem{DME2000}
Drzewi\ifmmode~\acute{n}\else \'{n}\fi{}ski A, Macio\l{}ek A and Evans R 2000 {\em Phys. Rev. Lett.\/} {\bf 85}(15) 3079--3082

\bibitem{EM87}
Evans R and Marconi U~M~B 1987 {\em The Journal of Chemical Physics\/} {\bf 86} 7138

\bibitem{DMC2000}
Drzewi\ifmmode~\acute{n}\else \'{n}\fi{}ski A, Macio\l{}ek A and Ciach A 2000 {\em Phys. Rev. E\/} {\bf 61}(5) 5009--5018


\bibitem{V2014}
Vasilyev O~A 2014 {\em Phys. Rev. E\/} {\bf 90}(1) 012138

\end{thebibliography}
\end{document}